\documentclass[iop,apj]{emulateapj_tt}
\pdfoutput=1
\usepackage[breaklinks, colorlinks,
  urlcolor=blue, citecolor=blue, linkcolor=blue]{hyperref}
\usepackage{graphicx}
\usepackage{multirow}
\usepackage{color}
\usepackage{marginnote}
\reversemarginpar

\usepackage{fancyhdr}
\def \solar{_\odot}
\def \pow10#1{\times 10^{#1}}
\newcommand\blfootnote[1]{%
  \begingroup
  \renewcommand\thefootnote{}\footnote{#1}%
  \addtocounter{footnote}{-1}%
  \endgroup
}

\tighten

\begin{document}

\submitted{Submitted to ApJ}

\title{Observations of environmental quenching in groups in the 11 Gyr since
  $\lowercase{z}=2.5$: different quenching for central and satellite galaxies}

\author{Tomer Tal\hyperlink{afflink}{$^{1}$}$^,$\hyperlink{afflink}{$^{13}$}}
\author{Avishai Dekel\hyperlink{afflink}{$^2$}}
\author{Pascal Oesch\hyperlink{afflink}{$^3$}}
\author{Adam Muzzin\hyperlink{afflink}{$^4$}}

\author{Gabriel B. Brammer\hyperlink{afflink}{$^5$}}
\author{Pieter G. van Dokkum\hyperlink{afflink}{$^3$}}
\author{Marijn Franx\hyperlink{afflink}{$^4$}}
\author{Garth D. Illingworth\hyperlink{afflink}{$^1$}}
\author{Joel Leja\hyperlink{afflink}{$^3$}}
\author{Daniel Magee\hyperlink{afflink}{$^1$}}
\author{Danilo Marchesini\hyperlink{afflink}{$^6$}}
\author{Ivelina Momcheva\hyperlink{afflink}{$^3$}}
\author{Erica J. Nelson\hyperlink{afflink}{$^3$}}
\author{Shannon G. Patel\hyperlink{afflink}{$^7$}}
\author{Ryan F. Quadri\hyperlink{afflink}{$^{7}$}}
\author{Hans-Walter Rix\hyperlink{afflink}{$^8$}}
\author{Rosalind E. Skelton\hyperlink{afflink}{$^{9}$}}
\author{David A. Wake\hyperlink{afflink}{$^{10,11}$}}
\author{Katherine E. Whitaker\hyperlink{afflink}{$^{12}$}}

\affiliation{\hypertarget{afflink}{$^1$ UCO/Lick Observatory, University of 
    California, Santa Cruz, CA 95064, USA;
    \href{mailto:tomer.tal@yale.edu}{tal@ucolick.org}}\\
  \hypertarget{afflink}{$^2$ Racah Institute of Physics,
    The Hebrew University, Jerusalem 91904, Israel}\\
  \hypertarget{afflink}{$^3$ Yale University Astronomy Department, 
    P.O. Box 208101, New Haven, CT 06520-8101 USA}\\
  \hypertarget{afflink}{$^4$ Leiden Observatory, Leiden University, 
    NL-2300 RA Leiden, The Netherlands}\\
  \hypertarget{afflink}{$^5$Space Telescope Science Institute, 
    3700 San Martin Drive, Baltimore, MD 21218, USA}\\
  \hypertarget{afflink}{$^6$Department of Physics and Astronomy, Tufts 
    University, Medford, MA 02155, USA}\\
  \hypertarget{afflink}{$^7$Carnegie Observatories, Pasadena, CA 91101, USA}\\
  \hypertarget{afflink}{$^8$Max-Planck-Institut f\"{u}r Astronomie, 
    K\"{o}nigstuhl 17, 69117 Heidelberg, Germany}\\
  \hypertarget{afflink}{$^{9}$South African Astronomical Observatory, 
    Observatory Road, Cape Town, South Africa}\\
  \hypertarget{afflink}{$^{10}$Department of Astronomy, University of 
    Wisconsin-Madison, Madison, WI 53706, USA}\\
  \hypertarget{afflink}{$^{11}$Department of Physical Sciences, The Open
    University, Milton Keynes MK7 6AA, UK}\\
  \hypertarget{afflink}{$^{12}$Astrophysics Science Division, Goddard 
    Space Flight Center, Greenbelt, MD 20771, USA}}

\begin{abstract}
  We present direct observational evidence for star formation quenching in 
  galaxy groups in the redshift range $0<z<2.5$.
  We utilize a large sample of nearly 6000 groups, selected by fixed cumulative
  number density from three photometric catalogs, to follow the evolving 
  quiescent fractions of central and satellite galaxies over roughly 11 Gyr.
  At $z\sim0$, central galaxies in our sample range in stellar mass from 
  Milky Way/M31 analogs ($M_\star/M\solar=6.5\pow10{10}$) to nearby massive 
  ellipticals ($M_\star/M\solar=1.5\pow10{11}$).
  Satellite galaxies in the same groups reach masses as low as twice 
  that of the Large Magellanic Cloud ($M_\star/M\solar=6.5\pow10{9}$).
  Using statistical background subtraction, we measure the average rest-frame 
  colors of galaxies in our groups and calculate the evolving quiescent 
  fractions of centrals and satellites over seven redshift bins.
  Our analysis shows clear evidence for star formation quenching in group
  halos, with a different quenching onset for centrals and their satellite 
  galaxies.
%
%
  Using halo mass estimates for our central galaxies, we find that star 
  formation shuts off in centrals when typical halo masses reach between 
  $10^{12}$ and $10^{13}M\solar$, consistent with predictions from the halo 
  quenching model.
  In contrast, satellite galaxies in the same groups most likely undergo 
  quenching by environmental processes, whose onset is delayed with respect 
  to their central galaxy.
  Although star formation is suppressed in all galaxies over time, the
  processes that govern quenching are different for centrals and satellites.
  While mass plays an important role in determining the star formation activity
  of central galaxies, quenching in satellite galaxies is dominated by
  the environment in which they reside.
%


\end{abstract}



\section{Introduction}
\label{intro}
 \blfootnote{\hypertarget{afflink}{
     \hspace{-0.5cm}\noindent\rule{5.5cm}{0.4pt}\vspace{0.05cm}\\
     $^{13}$ NSF Astronomy and Astrophysics Postdoctoral Fellow
     \marginnote{\includegraphics[width=1.0cm]{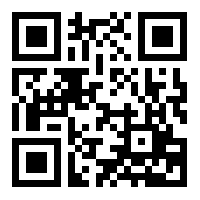}}[0.4cm]}}
 Star formation activity in galaxies at low redshift varies strongly with 
 local environment.
 Correlations between local galaxy density and star formation tracers 
 (e.g., star formation rates, stellar colors, morphology) have long been 
 observed in the nearby universe 
 (e.g., Oemler \citeyear{oemler_systematic_1974}; 
 Dressler \citeyear{dressler_galaxy_1980};
 Kauffmann et al. \citeyear{kauffmann_environmental_2004};
 Hogg et al. \citeyear{hogg_dependence_2004}; 
 Blanton et al. \citeyear{blanton_relationship_2005}; 
 Thomas et al. \citeyear{thomas_epochs_2005}).
 Many physical mechanisms have been shown to be relevant for regulating star
 formation in dense environments by driving cold gas away from galaxies 
 and by heating it up.
 The high fraction of non star forming galaxies in clusters and 
 massive groups implies that such mechanisms have already acted to quench 
 star formation in low redshift halos.

 Tremendous effort has been devoted in recent years to analyses of
 galaxy groups and lower mass environments.
 While such studies find that the properties of galaxies in groups are indeed 
 different than those of field galaxies (e.g.,
 Zabludoff \& Mulchaey \citeyear{zabludoff_properties_1998};
 Balogh et al. \citeyear{balogh_galaxy_2004}, \citeyear{balogh_colour_2009};
 Wilman et al. \citeyear{wilman_galaxy_2005};
 Weinmann et al. \citeyear{weinmann_properties_2006};
 Cooper et al. \citeyear{cooper_deep2_2007};
 Tran et al. \citeyear{tran_spectroscopically_2009};
 Patel et al. \citeyear{patel_dependence_2009}, 
 \citeyear{patel_star-formation-rate-density_2011};
 Bolzonella et al. \citeyear{bolzonella_tracking_2010};
 Peng et al. \citeyear{peng_mass_2010};
 Kova\v{c} et al. \citeyear{kovac_10k_2010};
 McGee et al. \citeyear{mcgee_dawn_2011};
 Quadri et al. \citeyear{quadri_tracing_2012};
 Muzzin et al. \citeyear{muzzin_gemini_2012};
 Geha et al. \citeyear{geha_stellar_2012};
 Rasmussen et al. \citeyear{rasmussen_suppression_2012};
 Hou et al. \citeyear{hou_group_2013}),
 it appears that environmental processes do not influence all galaxies equally
 (e.g.,
 Yang et al. \citeyear{yang_galaxy_2005};
 van den Bosch et al. \citeyear{van_den_bosch_importance_2008};
 Font et al. \citeyear{font_colours_2008};
 Skibba \& Sheth \citeyear{skibba_halo_2009};
 Skibba \citeyear{skibba_central_2009};
 Guo et al. \citeyear{guo_structural_2009};
 Peng et al. \citeyear{peng_mass_2012};
 Wetzel et al. \citeyear{wetzel_galaxy_2012}, \citeyear{wetzel_galaxy_2013};
 Knobel et al. \citeyear{knobel_colors_2013};
 Woo et al. \citeyear{woo_dependence_2013};
 Carollo et al. \citeyear{carollo_zurich_2013}).
 
 Central galaxies, defined as the most massive galaxies in their group or 
 cluster, are affected by processes whose star formation quenching efficiency 
 is proportional to the mass of the surrounding host halo.
 Such processes can act on the central galaxy itself (e.g., active 
 galactic nuclei: Kauffmann et al. \citeyear{kauffmann_unified_2000}; 
 Springel et al. \citeyear{springel_black_2005}; 
 Hopkins et al. \citeyear{hopkins_unified_2006}; 
 Fabian et al. \citeyear{fabian_observational_2012}) 
 or they can influence galaxies on halo scales (e.g., halo-quenching: 
 Birnboim \& Dekel \citeyear{birnboim_virial_2003};
 Dekel \& Birnboim \citeyear{dekel_galaxy_2006}).
 In addition to being affected by the same processes, satellite galaxies in 
 groups and clusters are also subject to gas and stellar stripping, 
 strangulation and harassment, all of which contribute to a gradual shut off 
 of star formation (e.g., 
 Gunn \& Gott \citeyear{gunn_infall_1972};
 Larson et al. \citeyear{larson_evolution_1980};
 Farouki \& Shapiro \citeyear{farouki_computer_1981};
 Byrd \& Valtonen \citeyear{byrd_tidal_1990};
 Moore et al. \citeyear{moore_galaxy_1996};
 Balogh et al. \citeyear{balogh_origin_2000}).
 Such interactions take place between satellite galaxies or between satellites 
 and the host halo itself, and are sensitive to the varying local gas and 
 galaxy density in the halo.

 Internal processes are likely also important drivers of star formation 
 quenching in all galaxies.
 Recent studies find that structural properties of galaxies, such as 
 central mass density and bulge-to-total mass ratio, are correlated
 with star formation activity at low and high redshift
 (e.g., Cheung et al. \citeyear{cheung_dependence_2012};
 Fang et al. \citeyear{fang_link_2013}; 
 Barro et al. \citeyear{barro_candels:_2013}).
 In addition, stellar velocity dispersion was found to be a good proxy for 
 galaxy color and star formation activity, essentially linking the inner 
 stellar body with the host dark matter halo (e.g., 
 Franx et al. \citeyear{franx_structure_2008};
 Wake et al. \citeyear{wake_revealing_2012};
 Bezanson et al. \citeyear{bezanson_evolution_2012};
 Weinmann et al. \citeyear{weinmann_velocity_2013}). 
 Such correlations indicate that the local properties of galaxies pose  
 essential conditions for star formation suppression.
  
 Here we perform a statistical analysis of galaxies in groups at $0<z<2.5$.
 Since groups assemble and evolve later than massive clusters, quenching
 processes are still important even at relatively late times.
 We are able to explore the average properties of central and satellite 
 galaxies over a large redshift range using only photometric catalogs and 
 statistical background subtraction.
 This approach is useful for deriving the evolution of galaxy properties 
 without the need of spectra (e.g., 
 Masjedi et al. \citeyear{masjedi_very_2006}, \citeyear{masjedi_growth_2008};
 Watson et al. \citeyear{watson_extreme_2012};
 Tal et al. \citeyear{tal_mass_2012}, \citeyear{tal_galaxy_2013};
 Wang \& White \citeyear{wang_satellite_2012};
 M\'{a}rmol-Queralt\'{o} et al. \citeyear{marmol-queralto_satellites_2012}, 
 \citeyear{marmol-queralto_characterizing_2013}).
 This may currently be the most efficient way to study galaxy 
 group halos even at moderate redshift. 

 Throughout the paper we adopt the following cosmological parameters:
 $\Omega_m=0.3$, $\Omega_{\Lambda}=0.7$ and $H_0=70$ km s$^{-1}$
 Mpc$^{-1}$.

  \begin{figure}
    \includegraphics[width=0.48\textwidth]{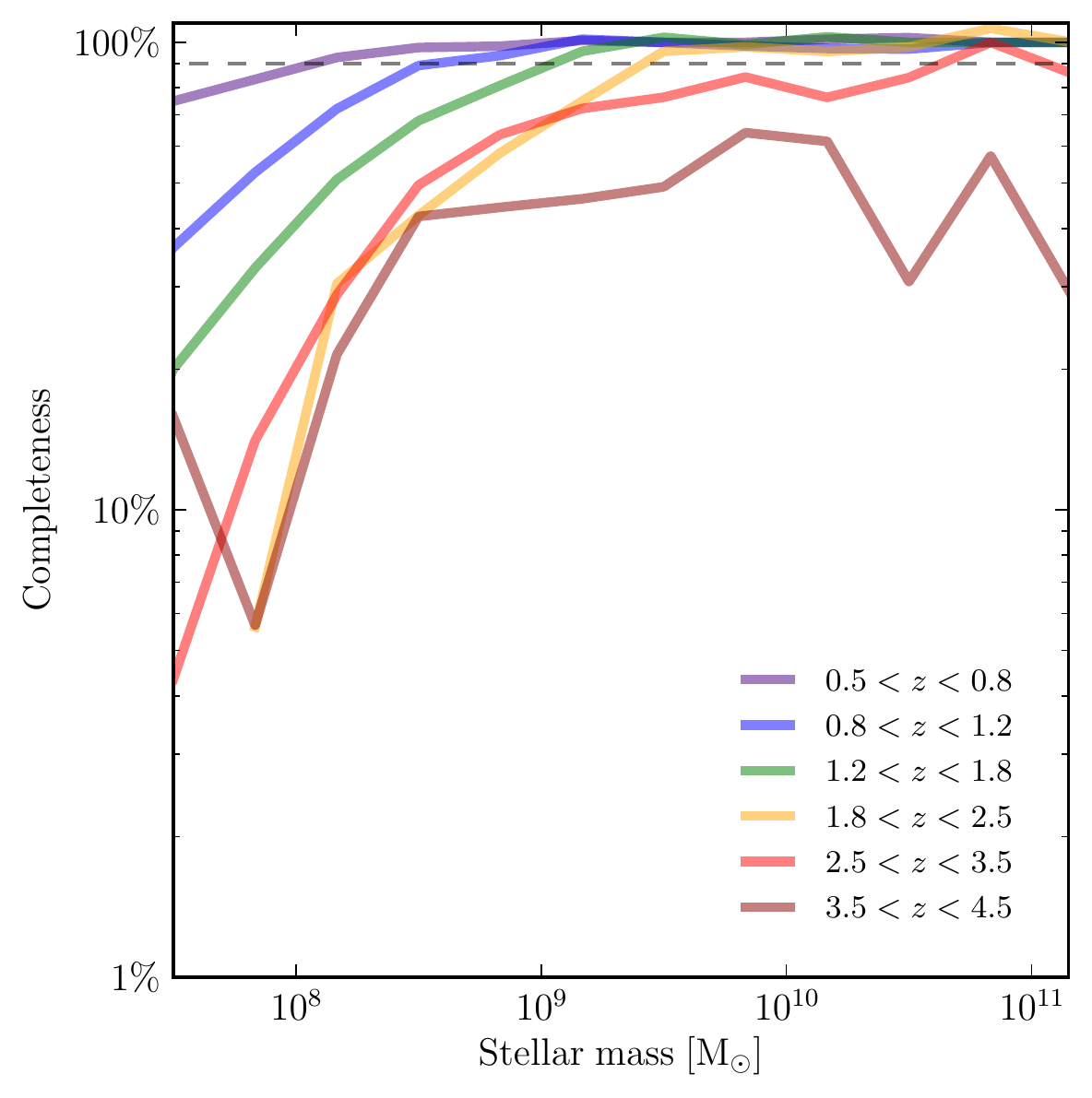}
    \caption{Completeness in the 3D-HST/CANDELS catalog as a function of 
      galaxy stellar mass.
      Completeness was calculated by comparing object detection in CANDELS/deep 
      with a re-combined subset of the exposures which reach the depth
      of CANDELS/wide (see Section \ref{sec:completeness} for details).
      Each of the curves represents this measurement in a different redshift
      bin.
      The catalogs are 90\% complete down to $1.1\pow10{9}M\solar$ at 
      $z<1.8$ and down to $2.6\pow10{9}M\solar$ at $z<2.5$.
    }
    \label{fig:comp}
  \end{figure}
  
\section{Photometric catalogs and Sample Selection}
\label{sec:data}
 Galaxies for this study were selected at $0<z<2.5$ from three datasets, 
 utilizing data from four surveys:
 UltraVISTA (McCracken et al. \citeyear{mccracken_ultravista:_2012}),
 3D-HST (Brammer et al. \citeyear{brammer_3d-hst:_2012}),
 Cosmic Assembly Near-infrared Deep Extragalactic Legacy Survey
 (CANDELS; Grogin et al. \citeyear{grogin_candels:_2011}),
 and Sloan Digital Sky Survey (SDSS; York et al. \citeyear{york_sloan_2000}).
 
 Low redshift galaxies were selected at $0.02<z<0.04$ from the NYU Value-Added 
 Galaxy Catalog (NYU-VAG; Blanton et al. \citeyear{blanton_new_2005}), which
 was derived using imaging and spectroscopic data from the seventh data release
 of SDSS (Abazajian et al. \citeyear{abazajian_seventh_2009}) and imaging data 
 from the Two Micron All Sky Survey 
 (Skrutskie et al. \citeyear{skrutskie_two_2006}).
 Redshifts were determined spectroscopically as part of SDSS for roughly 
 94\% of the galaxies in that catalog while the rest of the galaxies were 
 assigned the same redshift measurement as their nearest spectroscopic 
 neighbor.
 All galaxies in our low redshift sample are more massive than 
 $5\pow10{9}M\solar$, the stellar mass completeness limit at $z=0.04$ 
 (e.g., Wake et al. \citeyear{wake_revealing_2012}).
 
 Galaxies in the redshift range $0.2<z<1.2$ were identified from a public
 $K_s$-selected catalog (Muzzin et al. \citeyear{muzzin_public_2013})
 based on the first data release of UltraVISTA, an ongoing ultra deep 
 near-infrared survey with the European Southern Observatory VISTA 
 survey telescope.
 The catalog covers a total area of 1.62 deg$^2$ in the COSMOS field.
 It includes photometry in 30 bands and provides excellent photometric 
 redshifts ($\sigma_z/(1+z)=0.013$).
 Stellar mass completeness analysis was performed for this data set by 
 Muzzin et al. (\citeyear{muzzin_evolution_2013}), who found a 95\%
 completeness limit of $5\pow10{9}M\solar$ at $z<1.2$.
 
 Galaxies at $z>1.2$ were observed as part of the complimentary 3D-HST and 
 CANDELS surveys.
 These are near-infrared spectroscopic and imaging surveys with the Wide 
 Field Camera 3 on board the Hubble Space Telescope.
 Images and grism spectra are available from the surveys in five fields
 (AEGIS, COSMOS, GOODS-S, GOODS-N and UDS), covering a total area of roughly
 0.2 deg$^2$.
 Photometric catalogs, including excellent photometric redshifts 
 ($\sigma_z/(1+z)\sim0.02$) and stellar population synthesis model fits were 
 derived by the 3D-HST team using images from both surveys in addition to a 
 large set of ancillary data (R. Skelton 2014, in preparation).
 We determine the stellar mass completeness limits of the 3D-HST/CANDELS 
 data set in the following section.
  
 \begin{figure}
   \includegraphics[width=0.48\textwidth]{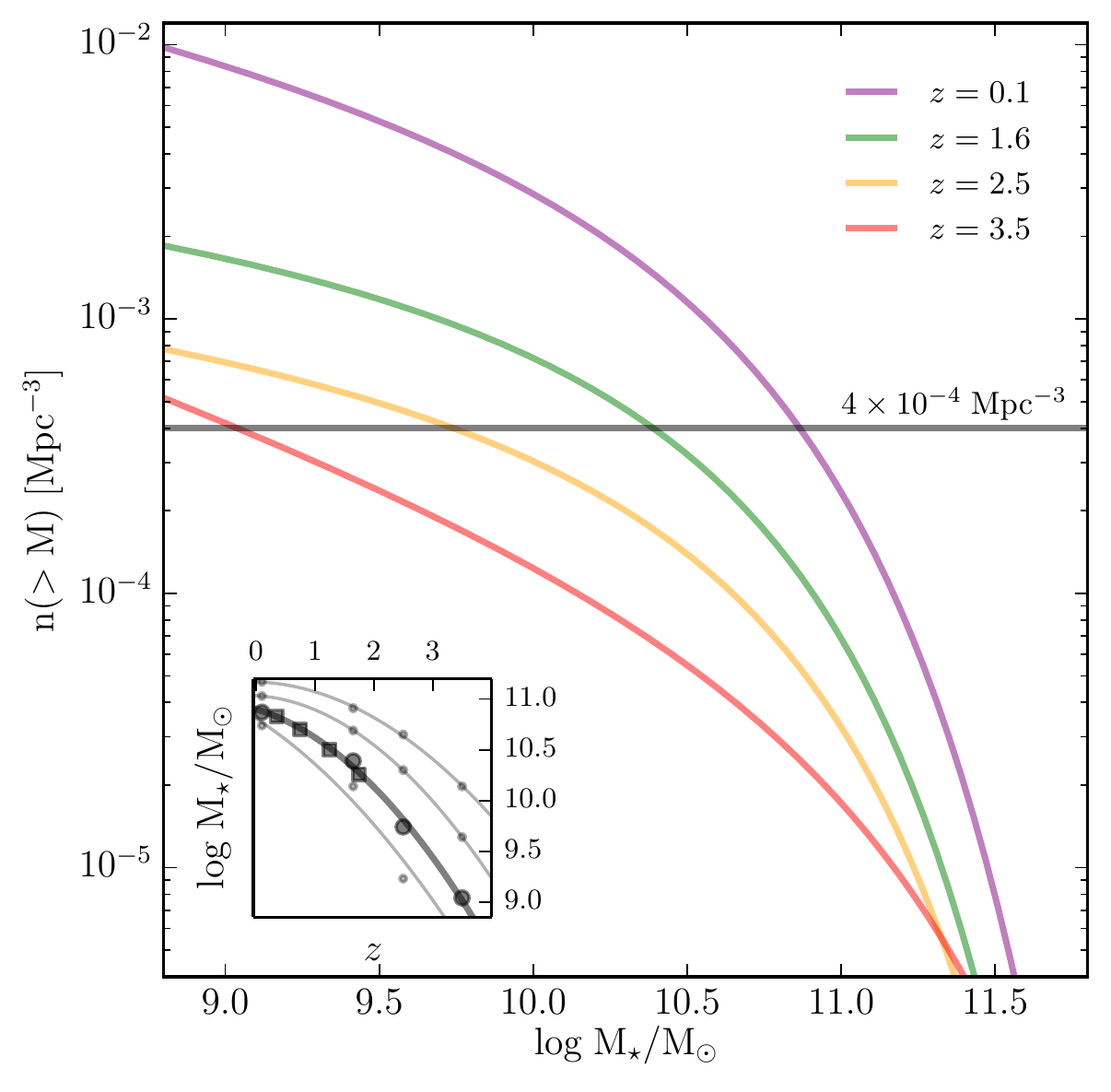}
   \caption{Sample matching across different redshift bins based on constant
     number density. 
     The color curves represent the cumulative stellar mass 
     functions from Marchesini et al. (\citeyear{marchesini_evolution_2009}).
     We calculated galaxy masses at four constant cumulative number density 
     values and derived mass-redshift relations which we fitted with 
     second order polynomials (inset figure). 
     For comparison, the square data points in the inset figure show the same 
     measurement using the stellar mass functions of Muzzin et al 
     \citeyear{muzzin_evolution_2013}.
     The selection at $n=4\pow10{-4}$ Mpc$^{-3}$ is marked for reference
     (gray horizontal line and thick curve in the inset figure).
     We utilize the resulting mass-redshift relations to select central galaxies
     at $0<z<2.5$.
   }
   \label{fig:m09_sel}
 \end{figure}

 \subsection{Completeness in 3D-HST/CANDELS}
 \label{sec:completeness}
  We estimate the completeness in stellar mass of the 3D-HST/CANDELS v3.0 
  catalog by comparing source detection in the wide and deep components of the 
  CANDELS survey.
  We do so by creating a detection image over the CANDELS/deep area with 
  the same exposure time and depth as the shallower CANDELS/wide survey.
  Although CANDELS/deep covers only about 17\% of the total survey area, it
  reaches roughly twice the depth of CANDELS/wide (with approximately four 
  times the exposure time), thus making it suitable for calculating the 
  fraction of missed objects as a function of stellar mass.

  Following the same methodology that was used to create the science grade 
  images, we reduced only 25\% of the data that is available in the
  GOODS-S/deep field to match the depth of the wide component.
  From CANDELS/deep we selected observations from visits 1 and 2 (as defined
  in Grogin et al. \citeyear{grogin_candels:_2011}) and combined 
  them to create 2-orbit depth images in F125W and F160W.
  From 3D-HST we combined all F140W direct images in GOODS-S into a
  single orbit depth image.
  Finally, we created a noise equalized detection image following the same 
  procedure that was used to produce the 3D-HST/CANDELS catalog.
  We multiplied each of the three science grade images (sci$_i$) by the 
  square-root of their corresponding exposure-time weight-mask 
  ($\sqrt{\mathrm{weight}_i}$) and averaged them according to the following 
  equation:
  \begin{equation}
    \label{eq:mass-z}
    Detection\ image = \frac{\sum{\left(\sqrt{\mathrm{weight}_i}\times 
        \mathrm{sci}_i\right)}}
    {\sum{\sqrt{\mathrm{weight}_i}}}
  \end{equation}

  Source identification and matching was then performed on both depth level 
  images.
  We ran SExtractor (Bertin \& Arnouts \citeyear{bertin_sextractor:_1996})
  on the newly combined ``shallow'' image as well as on the deep GOODS-S 
  image using the same parameters that were utilized to create the photometric 
  catalog.
  Figure \ref{fig:comp} shows the fraction of detected objects in 
  the shallow image compared to the deep image as a function of mass and
  redshift.
  Input redshift and stellar mass estimates were taken from the 3D-HST/CANDELS 
  catalog, for which detected galaxies from both images were matched to within 
  3 pixels, the typical size of the WFC3 point spread function.
  On that basis we estimate that the catalogs are 90\% complete in stellar 
  mass down to $\log{(M_s/M_\odot)} = 9.04$ at $z<1.8$ and they reach 
  $\log{(M_s/M_\odot)} = 9.41$ at $z<2.5$.
  At $z<3.5$ the catalogs are 75\% complete down to 
  $\log{(M_s/M_\odot)} = 9.14$.
  
 \subsection{Cumulative Number Density Matching}
 \label{sec:nselect}
  Several recent studies have demonstrated that selecting galaxy samples at 
  a fixed cumulative number density effectively allows tracing the evolution 
  of a given population of galaxies over different epochs
  (e.g., Gao et al. \citeyear{gao_early_2004}; 
  van Dokkum et al. \citeyear{van_dokkum_growth_2010};
  Papovich et al. \citeyear{papovich_rising_2011};
  Leja et al. \citeyear{leja_tracing_2013};
  Behroozi et al. \citeyear{behroozi_using_2013}).
  In this study we followed the same technique to match galaxy populations 
  over the redshift range $0.02<z<3.5$.
  We calculated the cumulative number density curves of the 
  redshift samples from Marchesini et al. (\citeyear{marchesini_evolution_2009})
  and measured the implied stellar mass evolution at four number
  density values: $n=1, 2, 4, 6\pow10{-4}$ Mpc$^{-3}$ 
  (Figure \ref{fig:m09_sel}).
  For example, the resulting mass-redshift relation at 
  $n=4\pow10{-4}$ Mpc$^{-3}$ is:
  \begin{equation}
    \log(M_\star/M\solar)=-0.10z^2-0.18z+10.90
    \label{eq1}
  \end{equation}
  Similar relations for the other number density samples are given in 
  Appendix \hyperref[sec:appendixa]{A}.
  Toy model based estimates of the mass evolution at $z>1$ imply
  a similar slope at high redshift (Dekel et al. \citeyear{dekel_toy_2013}).

  \begin{table}[t]
    \caption{Central galaxy selection criteria at  $n=4\pow10{-4}$ Mpc$^{-3}$}
    \centering
    \begin{tabular}{l c c c c}
      \hline\hline
      Data set & Redshift & M$_{med}$/M$\solar$ & R$_{v}$ [kpc] &
      N$_\mathrm{cen}$ \\
      \hline {\vspace{-5px}}\\
      {\vspace{5px}}SDSS & $0.02<z<0.04$ & 7.9$\pow10{10}$ & 250 & 251 \\
      \multirow{4}{*}{UltraVISTA} & $0.2<z<0.4$ & 6.9$\pow10{10}$ & 
        220 & 261 \\
      & $0.4<z<0.6$ & 6.1$\pow10{10}$ & 200 & 460 \\
      & $0.6<z<0.85$ & 5.2$\pow10{10}$ & 180 & 1080 \\
      {\vspace{5px}}& $0.85<z<1.2$ & 4.1$\pow10{10}$ & 150 & 2281\\
      \multirow{3}{*}{3D-HST} & $1.2<z<1.8$ & 2.5$\pow10{10}$ & 110
      & 707 \\
      & $1.8<z<2.5$ & 1.2$\pow10{10}$ & 75 & 915 \\
      {\vspace{3px}}& $2.5<z<3.5$ & 2.8$\pow10{9}$ & 40 & 3054\\
      \hline
      \label{tab:selection}
    \end{tabular}
  \end{table}

  We note that while the cumulative number density matching technique provides
  an estimate for the mass evolution of a given galaxy population over time, 
  it relies on the assumption that the rank order of galaxies does not change 
  dramatically over the studied redshift range.
  The presence of a significant major merger rate may change the mass ranking
  of galaxies and therefore render this assumption inadequate.
  In addition, a high major merging rate would affect this measurement by 
  directly changing the number of galaxies at a given mass.

 \subsection{Central Galaxy Identification}
  We divided all galaxies from the UltraVISTA and 3D-HST/CANDELS catalogs 
  into six redshift bins such that each bin spans roughly 1.5 to 2 Gyr.
  The time span of the SDSS redshift bin is shorter, as it was limited by the 
  catalog mass completeness limit.
  At each redshift we selected central galaxy candidates in $\log{M_\star}$ 
  bins of width 0.3 dex with an evolving median mass according to the above
  calculated mass-redshift relation (e.g., Equation \ref{eq1}).
  We considered a galaxy to be central if no other, more massive, galaxy was 
  found within two projected virial radii of the host halo of the central 
  candidate.
  Virial radius estimates for host halos around centrals at a given median 
  mass and a given redshift were measured from catalogs extracted from the 
  semi-analytic model of Guo et al. (\citeyear{guo_dwarf_2011}).
  Median stellar mass, redshift limits and virial radius estimates for 
  central galaxies in each of the bins at $n=4\pow10{-4}$ Mpc$^{-3}$ are given 
  in Table \ref{tab:selection}.
  At $z=0$, central galaxies in our sample reach 
  $M_\star/M\solar=6.5\pow10{10}$, roughly the mass of the Milky Way and M31.
  Detailed selection criteria for all samples are presented in Appendix 
  \hyperref[sec:appendixa]{A}.

 \subsection{Rest frame $UVJ$ Colors}
 \label{sec:uvj}
  Global star formation rates have been measured in distant galaxies using 
  numerous techniques, often relying on photometric estimates of rest frame
  colors.
  Williams et al. (\citeyear{williams_detection_2009}) showed 
  that galaxies out to $z=2$ can be reliably separated into either 
  ``star forming'' or ``quiescent'' based on a rest frame $U-V$ vs. 
  $V-J$ ($UVJ$) color selection.
  Recently, Muzzin et al. (\citeyear{muzzin_public_2013}) extended this 
  measurement to $z=3.5$ using redshift and rest frame color estimates from the 
  UltraVISTA survey catalog.

  Following the same approach, we identified each galaxy in our sample as 
  either star forming or quiescent based on the derived rest frame colors in 
  the catalogs.
  We adopted the $UVJ$ threshold values from Muzzin et al.
  (\citeyear{muzzin_public_2013}) to separate galaxies at $z>0.2$ 
  in the UltraVISTA and 3D-HST/CANDELS catalogs.
  At lower redshift we found equivalent thresholds in the $u-g$ vs. $g-J$ color 
  plane that we used for selecting galaxies from the NYU-VAG catalog.
  We note that the relatively poor $u-$band photometry in SDSS implies 
  that rest frame $u-g$ color estimates at low redshift are significantly more
  scattered than $U-V$ estimates in UltraVISTA and 3D-HST (see Appendix 
  \hyperref[sec:appendixb]{B}
  for details).

 \begin{figure}
   \includegraphics[width=0.48\textwidth]{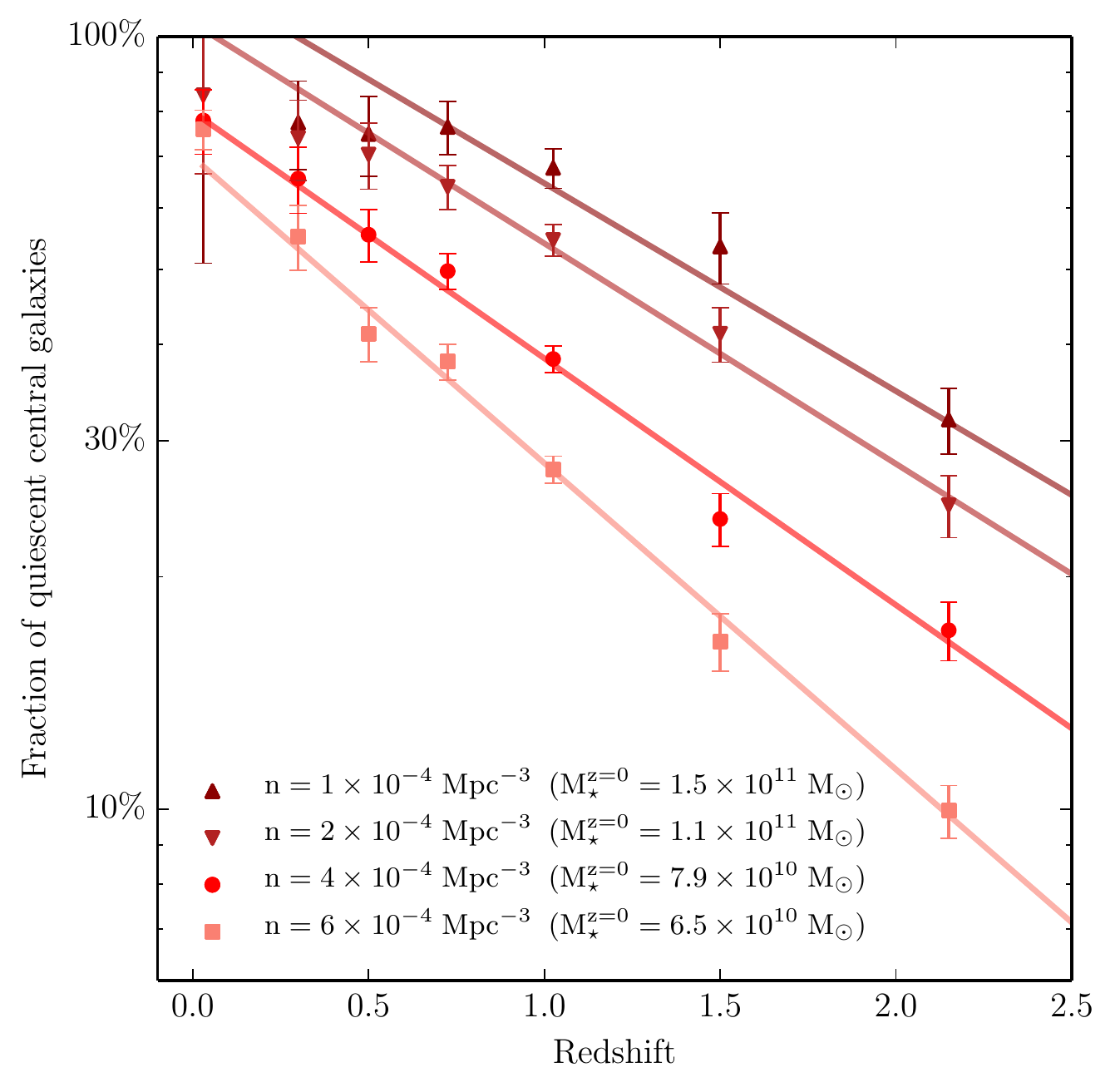}
   \caption{Redshift evolution of the fraction of quiescent central galaxies.
     Each set of points represent a fixed cumulative number density selection
     of central galaxies and are fitted by a line.
     Stellar mass estimates at $z=0$ are also noted for each value of $n$.
     Error bars are calculated as propagated $\sqrt{N}$ noise estimates.
     While the fraction of quiescent galaxies increases with time for all
     samples, lower number density samples (which correspond to earlier mass 
     growth) preferentially get quenched at higher redshift.
   }
   \label{fig:nsf_n}
 \end{figure}

\section{Halo quenching of central galaxies}
\label{sec:cenfrac}
 Star formation suppression in central galaxies is often linked to the 
 mass of their surrounding host halos.
 In this ``halo quenching'' model, strong shocks in halos above a
 critical mass $M_{\mathrm{crit}}\sim10^{12}M\solar$ heat up any infalling 
 cold gas and thus prevent further star formation in the central galaxy itself
 (e.g., Birnboim \& Dekel \citeyear{birnboim_virial_2003};
 Dekel \& Birnboim \citeyear{dekel_galaxy_2006}).
 This model predicts that the value of $M_{crit}$ is mostly constant at low
 redshift and that at high redshift significant cold gas inflows allow star 
 formation even in halos more massive than $M_{\mathrm{crit}}$ .
 Recently, Behroozi et al. (\citeyear{behroozi_average_2013}) matched 
 observed galaxies to simulated dark matter halos to show that star
 formation efficiency peaks in halos with a typical mass around 
 $5\pow10{11}M\solar$ and that it drops quickly in more massive halos.
 Here we follow an alternative approach and estimate the halo mass at which 
 quenching becomes important by analyzing the fraction of quiescent central 
 galaxies as a function of redshift and mass.

 \begin{figure}
   \includegraphics[width=0.48\textwidth]{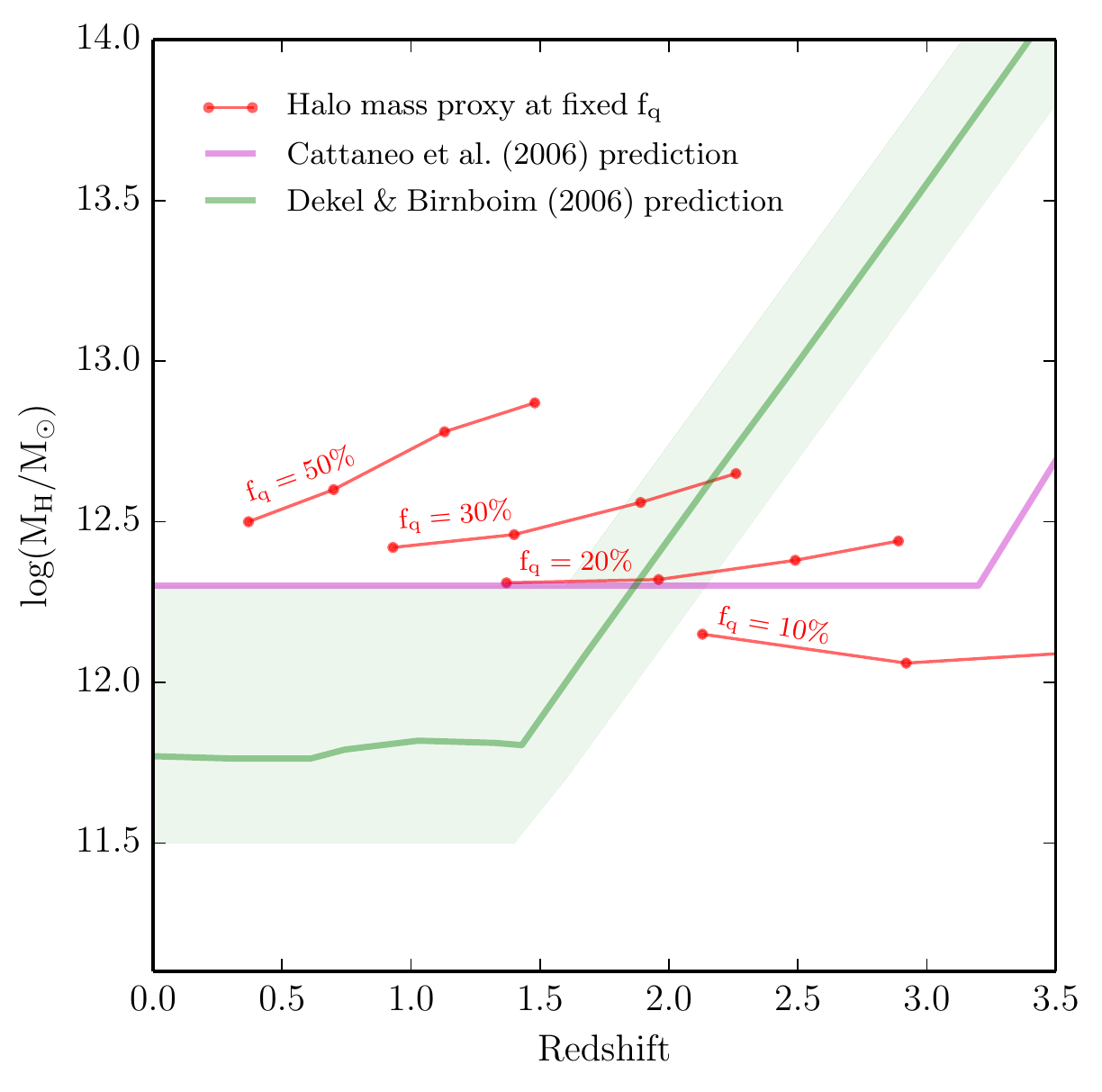}
   \caption{Predicted and observed redshift evolution of the threshold mass 
     for halo quenching.
     Halo mass proxy estimates are calculated under Press Schechter formalism 
     for each pair of stellar mass and number density values.
     The red points and lines represent the halo mass estimates and redshifts 
     of each number density sample when the fraction of quiescent galaxies 
     reaches 10\%, 20\%, 30\% and 50\%.
     Also plotted are theoretical predictions from the halo-quenching model, 
     as calculated by Cattaneo et al. (\citeyear{cattaneo_modelling_2006}, 
     magenta line) and by Dekel \& Birnboin (\citeyear{dekel_galaxy_2006}, 
     green line).
     The observed typical halo mass at which central galaxies are quenched is
     between $10^{12}M\solar$ and $10^{13}M\solar$, broadly in agreement with 
     predicted values.
   }
   \label{fig:mh_z}
 \end{figure}

\subsection{Early quenching of central galaxies}
  We use rest frame $UVJ$ color estimates to identify quenched central galaxies
  in each redshift sample.
  Figure \ref{fig:nsf_n} shows the redshift dependence of the fraction of
  quiescent central galaxies in each of the four number density selected 
  samples.
  The red circles and line represent the $n=4\pow10{-4}$ Mpc$^{-3}$ sample
  while lighter and darker points and lines show the same measurement at higher
  and lower fixed number density values.
  Also noted in the figure are $z=0$ stellar mass estimates of galaxies
  in each sample.
  The fraction of quiescent central galaxies at all number densities seem to 
  be increasing at nearly the same rate at $z\lesssim 2.5$.
  The evolving quiescent fractions that are presented in Figure \ref{fig:nsf_n}
  are consistent with results from recent similar analyses at fixed cumulative
  number density (Patel et al. \citeyear{patel_hst/wfc3_2013}; 
  Lundgren et al. \citeyear{lundgren_tracing_2013}).

  Figure \ref{fig:nsf_n} shows that central galaxy quenching occurs early.
  The quiescent fraction of the lowest number density sample 
  ($1.5\pow10{11}M\solar$ at $z=0$) reaches 30\% already at $z>2.0$.
  Lower mass central galaxies, drawn from the higher number density samples, 
  also begin quenching at relatively early times and are more than 10\% 
  quenched by $z=2.0$.

  This result agrees qualitatively with the halo quenching model which
  predicts that centrals cease forming stars promptly after their host halo
  reaches the threshold mass $M_{\mathrm{crit}}$.
  In the following Section, we estimate the evolution of this threshold mass 
  from observations.

 \subsection{Halo Mass and Quenching}
  We examine the influence of halo mass on star formation suppression in 
  central galaxies by estimating the typical host halo mass at 
  four fixed values of quenched fractions ($f_q$ = 10\%, 20\%, 30\% and 50\%).
  We start by using the line fits from Figure \ref{fig:nsf_n} to extract the 
  redshift at which each sample of central galaxies crosses the selected value
  of $f_q$.
  We then estimate the total stellar mass of central galaxies at this redshift 
  using the derived relation of each number density sample (e.g., Equation 
  \ref{eq1}).
  Finally, we estimate a proxy for the group host halo mass using the 
  total stellar mass and number density under Press-Schechter formalism 
  (Press \& Schechter \citeyear{press_formation_1974}).

  Figure \ref{fig:mh_z} shows the halo mass estimate of each number density 
  sample at each value of $f_q$ as a function of redshift (red points and 
  lines).
  Also shown are predictions from two theoretical investigations of the halo
  quenching model.
  The green line represents the estimates from Dekel \& Birnboim 
  (\citeyear{dekel_galaxy_2006}), who follow an analytic approach to study 
  the formation of strong shocks in spherically symmetric halos.
  In this analysis, the critical mass value was estimated at 
  $\log(M_{\mathrm{crit}}/M\solar)\sim11.7$ with significant uncertainty of 
  roughly half a dex.
  The magenta line is based on estimates from Cattaneo
  et al. (\citeyear{cattaneo_modelling_2006}), who fine-tune the value of
  $M_{\mathrm{crit}}$ in their semi-analytic simulations to best fit the 
  color-magnitude distribution of $z\sim0$ galaxies and the Lyman-break galaxy 
  luminosity function at $z\sim3$.
  These analyses use different tools to study the same model, according to 
  which galaxies that reside inside a host halo more massive than some critical 
  mass are efficiently quenched at low redshift (horizontal green and purple 
  lines in Figure \ref{fig:mh_z}).
  At high redshift, high infall rates of cold narrow streams allow star 
  formation in higher mass halos (diagonal lines).
  The upturn redshift in Dekel \& Birnboim (\citeyear{dekel_galaxy_2006}) is
  allowed in the range $1<z_{\mathrm{crit}}<3$.

 \begin{figure}
   \includegraphics[width=0.48\textwidth]{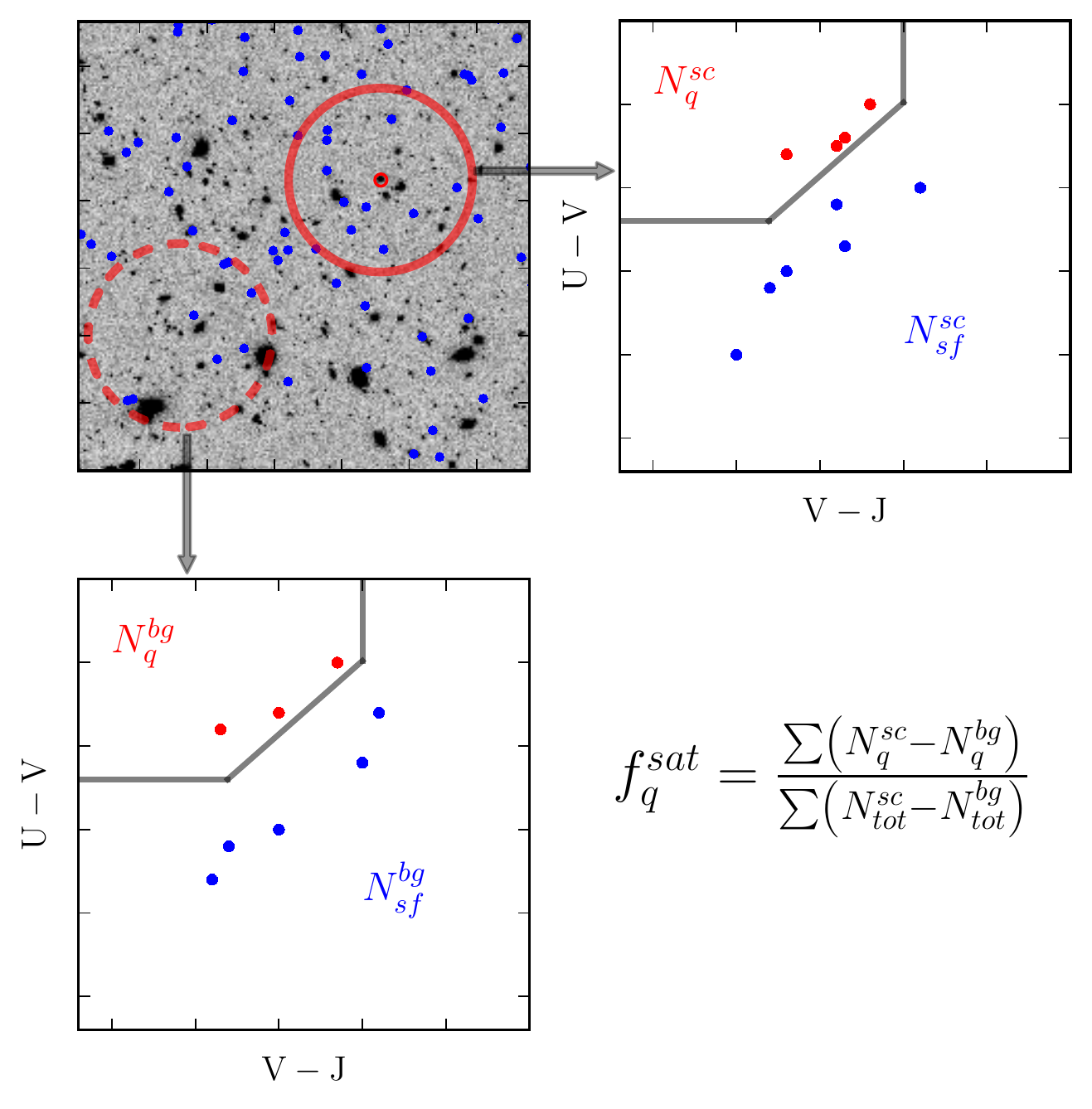}
   \caption{Demonstration of the technique used to derive quiescent satellite 
     galaxy fractions.
     Rest frame colors of satellite galaxy candidates were measured in apertures
     of size $r_{vir}$ around selected centrals (solid red circle) and used to 
     separate candidates into quiescent and star forming bins.
     A similar measurement was performed around randomly selected positions in 
     the imaging field (dashed red circle).
     The total number of quiescent sources in all background apertures 
     ($N_q^{bg}$) was subtracted from the total number of quiescent satellite 
     candidates ($N_q^{sc}$), and the result was divided by the difference 
     between all background and satellite candidate galaxies.
     The resulting fraction ($f_q^{sat}$) represents the average fraction of
     quiescent satellite galaxies around all centrals at a given redshift.
     While satellite galaxies cannot be identified individually around any one
     central, the ensemble of 6000 groups across all redshifts allows a 
     clear statistical separation between satellite and background 
     galaxies.
   }
   \label{fig:examp}
 \end{figure}

  Theoretical investigations of the halo quenching model agree with the 
  observed result in that there seems to exist a threshold host halo mass above 
  which star formation in central galaxies is suppressed efficiently.
  The implied typical range of the threshold halo masses
  $10^{12}\lesssim M_{\mathrm{crit}}/M\solar\lesssim 10^{13}$
  is consistent with the values favored by Cattaneo et al. 
  (\citeyear{cattaneo_modelling_2006}) and by Dekel \& Birnboim 
  (\citeyear{dekel_galaxy_2006}).
  We note that in addition to halo mass, local properties of the stellar
  component of galaxies (as traced by, e.g., central mass density and velocity 
  dispersion) likely help determine star formation activity in these galaxies.

\section{Environmental quenching of satellite galaxies}
\label{sec:sats}
 Measurements of the properties of individual satellites require that galaxies 
 are uniquely associated to their parent halo, often relying on extensive
 spectroscopic data sets.
 Here we follow an alternative approach and perform a statistical analysis 
 of the fraction of quiescent satellites around identified central galaxies.
 Using photometric redshift measurements we quantify and subtract the 
 contribution of foreground and background sources and derive the evolving
 quiescent satellite fraction in halos binned by mass and redshift.
 For the bulk of this analysis we utilize the $n=4\pow10{-4}$ Mpc$^{-3}$ sample 
 in order to maximize completeness, redshift and source density.
 At this value of $n$, galaxies are at least 10 times more massive 
 than the 90\% completeness limit at $z<1.8$, allowing us to study the 
 satellites of a mass evolving sample of central galaxies with a 
 satellite-central mass ratio $\geq$1:10.
 At higher values of $n$ completeness reaches 90\% at lower redshift while at 
 lower $n$ the density of satellite galaxies is insufficiently high for 
 background subtraction.

 \subsection{Average Quiescent Fractions}
 \label{sec:av_fq}
  We select all satellite galaxy candidates from the UltraVISTA 
  and 3D-HST/CANDELS catalogs within the virial radius of each central galaxy 
  and within $dz\leq0.05$ and $M_{\mathrm{satellite}}/M_{\mathrm{central}}\geq 0.1$.
  Galaxies in SDSS were selected from the NYU-VAG catalog within $dz\leq 0.01$ 
  of their central galaxies.
  Virial radius estimates at all redshifts were acquired from the Guo et al. 
  (\citeyear{guo_dwarf_2011}) semi-analytic model.
  Redshift thresholds were defined based on the redshift accuracy of the 
  catalogs and mass limits were determined to maximize the number of detected 
  satellites according to the calculated completeness value in stellar mass.

  We measure the rest frame $UVJ$ color ($ugJ$ in SDSS) of each candidate 
  satellite and count the number of quiescent and star forming galaxies.
  We then perform an identical measurement in ten randomly positioned 
  apertures per galaxy to get an estimate of background and foreground source
  contamination.
  Finally, we subtract the average number of contaminating sources in each 
  selection category (quiescent or star forming) from the number of candidate 
  satellites in the same category.
  From the resulting distribution of $UVJ$ and $ugJ$ values we calculate the
  average fraction of quiescent satellite galaxies at each redshift.
  Figure \ref{fig:examp} shows a demonstration of this procedure.
  We also measured the fraction of quiescent background and foreground
  galaxies and the standard deviation of values in the ten stacks of randomly 
  positioned apertures.

 \begin{figure}
   \includegraphics[width=0.48\textwidth]{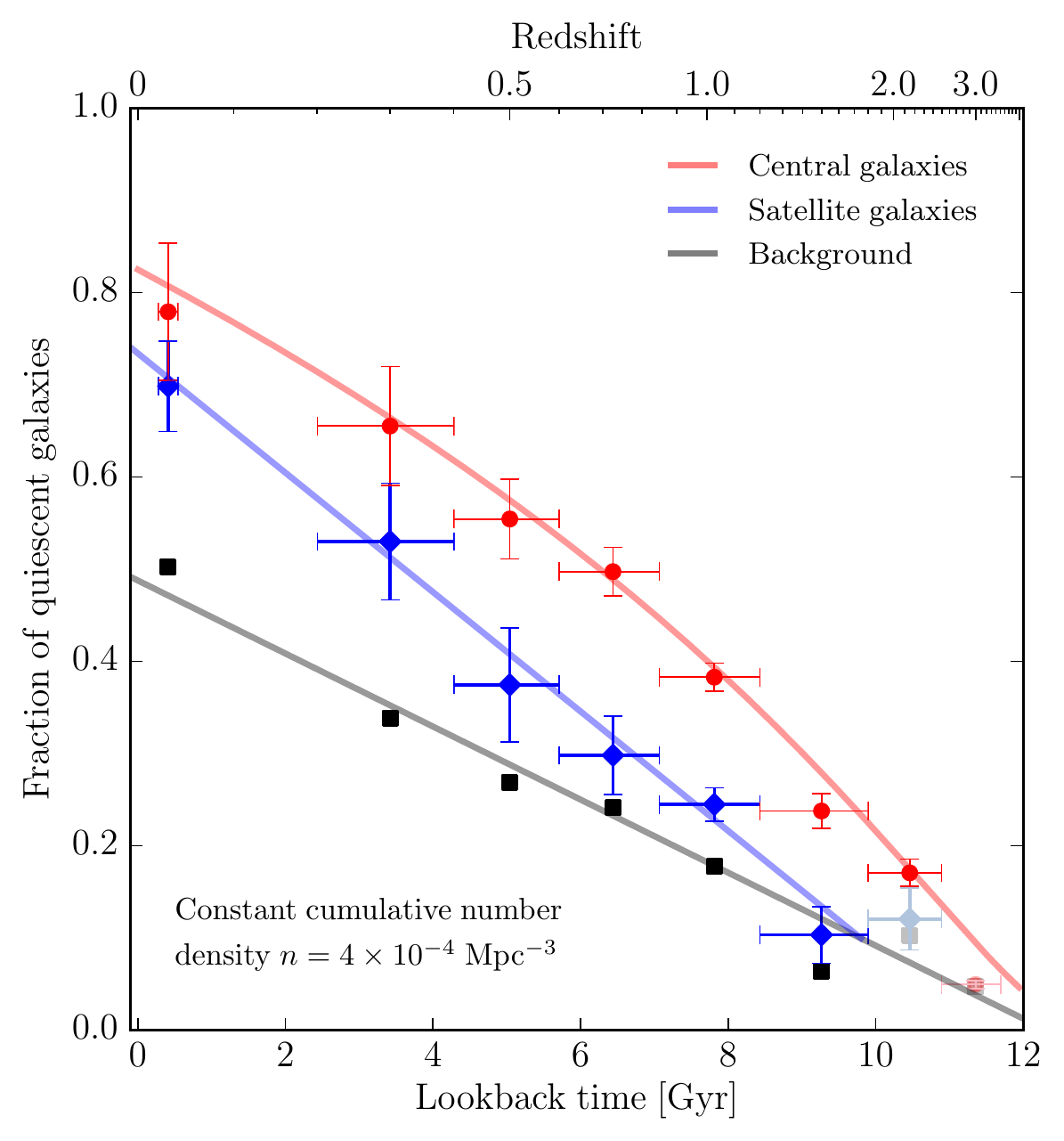}
   \caption{Environmental quenching of satellite galaxies.
     Red points and blue diamonds represent the fraction of quiescent 
     central and satellite galaxies, respectively.
     Black squares show the fraction of quiescent galaxies which are at the 
     same redshift and mass as the satellites but are selected from randomly
     positioned apertures (``background'' sources).
     The blue and black lines are fits to the data and the red line is the same
     as in Figure \ref{fig:nsf_n}.
     Pale points represent the same measurements at stellar masses below the
     90\% completeness limit.
     Although all galaxies get quenched over time, central, satellite and 
     background galaxies do so at different times and rates.
     While central galaxies are quenched early, star formation suppression in 
     their satellites is delayed by a few Gyr.
     In addition, the quenching rate of satellites is faster than that of
     background galaxies at the same mass.
     The observed quenching rate and onset of satellite galaxies imply that 
     star formation in these galaxies is strongly affected by environmental 
     processes.
   }
   \label{fig:nsf_frac}
 \end{figure}

  Figure \ref{fig:nsf_frac} shows the fraction of quiescent galaxies 
  as a function of lookback time and redshift at $n=4\pow10{-4}$ Mpc$^{-3}$.
  Red data points represent central galaxies, black points show
  galaxies in randomly positioned apertures and blue points show the 
  fraction of quiescent satellite galaxies.
  The pale pink data point at $z=3$ and the gray and light blue data points 
  at $z>2$ reflect the same measurement where the 3D-HST/CANDELS catalog is 
  less than 90\% complete in stellar mass.
  The plotted blue error bars represent the statistical measurement error and 
  are equal to the standard deviation of measured values from the 10 random
  apertures in each redshift bin.
  The red error bars are the propagated $\sqrt{N}$ noise estimates of the
  central quiescent fraction measurement.
  The black and blue solid lines show a linear fit to the data at $z<2$ and 
  the red line shows the line fit to the central galaxy points in 
  redshift-log($f_q$) space from Figure \ref{fig:nsf_n}.
  
 \subsection{Star Formation Quenching in Satellite Galaxies}
 \label{sec:sat_quench}
  The most striking result that is evident from Figure \ref{fig:nsf_frac} is the
  increase in the fraction of quiescent satellite galaxies compared
  to the overall background population.
  While on average all galaxies become more quiescent with time, 
  they do so at different epochs.
  Central galaxies, which are selected to be more massive than their satellites,
  increase their quiescent fraction at early times.
  Satellites and background galaxies exhibit a mostly 
  constant rate of increase in their quiescent fraction since $z\sim2$.
  At that redshift, it is not possible to separate satellites from the general 
  population of similar mass galaxies solely from their star formation activity.
  As time progresses, the quiescent fraction of satellite galaxies 
  increases compared to background galaxies and by $z\sim0$, the fraction of 
  non star forming satellite galaxies is nearly as high as that of their 
  centrals.

  Satellite and background galaxies in our sample were selected to be 
  identical in stellar mass and redshift, implying that the different quiescent 
  fractions at $z<2$ are likely attributable to the environment in which 
  these galaxies reside.
  In addition to halo quenching, which is expected to influence all galaxies, 
  satellite galaxies in groups are subject to processes such as gas and 
  stellar stripping, harassment and strangulation, all of which may assist in 
  suppressing star formation.
  Moreover, satellite orbits in the host halo are typically eccentric (e.g., 
  van den Bosch \citeyear{van_den_bosch_substructure_1999}) and as a 
  consequence they spend a large fraction of the time outside of the virial 
  radius.
  This implies that the quenching time scale of each satellite galaxy depends
  on its orbit, as well as on any interactions that the galaxy experiences. 
  Therefore, environmental processes may act rapidly to suppress star
  formation in some satellite galaxies or may take a long time to quench other
  galaxies.
  Nevertheless, the average quiescent fraction of the ensemble of satellite 
  galaxies increases over time compared to that of background galaxies at the
  same mass.

 \subsection{Delayed Onset of Satellite Quenching}
 \label{sec:tdelay}
  Quenching of satellite galaxies in groups does not seem to begin
  immediately upon their infall into the group halo.
  Wetzel et al. (\citeyear{wetzel_galaxy_2012}) utilized galaxy group catalogs
  based on SDSS data in addition to a high resolution $N$-body simulation in 
  order to estimate the time delay between satellite galaxy infall and 
  quenching.
  They find that star formation in satellites continues unaffected for 2-4 Gyr 
  after first infall.
  When quenching begins, it takes place on a rapid time scale that depends on
  the stellar mass of the satellites.
  Wetzel et al. (\citeyear{wetzel_galaxy_2012}) conclude that satellite galaxy
  quenching is similar to that of central galaxies for 2-4 Gyr after being
  accreted into a group halo, and is significantly more rapid thereafter.

  Our results indeed show that satellite quenching is delayed in respect to 
  their centrals, although we cannot measure the delay time scale as the 
  current data hold no information regarding satellite accretion times.
  Instead, we estimate the initial quenching time delay by comparing the
  time difference between the onset of central and satellite galaxy quenching.
  To do so we measured the time difference between the central and satellite 
  galaxy curves at $f_q\sim0.1$ (horizontal distance between the red and blue 
  curves in Figure \ref{fig:nsf_frac}), when the effects of environmental 
  processes first separate the quiescent fraction of satellites from that of 
  background galaxies.
  We repeated the measurement for an additional number density selection, 
  $n=6\pow10{-4}$ Mpc$^{-3}$, by finding the best fit curves to the data in 
  the same way as discussed in Section \ref{sec:av_fq} and presented in 
  Figure \ref{fig:nsf_frac}.
  We note that selected satellites at this value of $n$ reach 
  $M_\star/M\solar=6.5\pow10{9}$ at $z=0$, roughly twice the stellar mass of 
  the Large Magellanic Cloud.

  The derived initial time delay is 1.5$\pm$0.3 Gyr and 2.2$\pm$0.5 Gyr for 
  the $n=4\pow10{-4}$ Mpc$^{-3}$ and $n=6\pow10{-4}$ Mpc$^{-3}$ samples, 
  respectively.
  Both values are slightly shorter than the estimate of 2-4 Gyr that was 
  calculated for similarly massive galaxies at $z=0$ from Wetzel et al. 
  (\citeyear{wetzel_galaxy_2012}).

 \begin{figure}
   \includegraphics[width=0.48\textwidth]{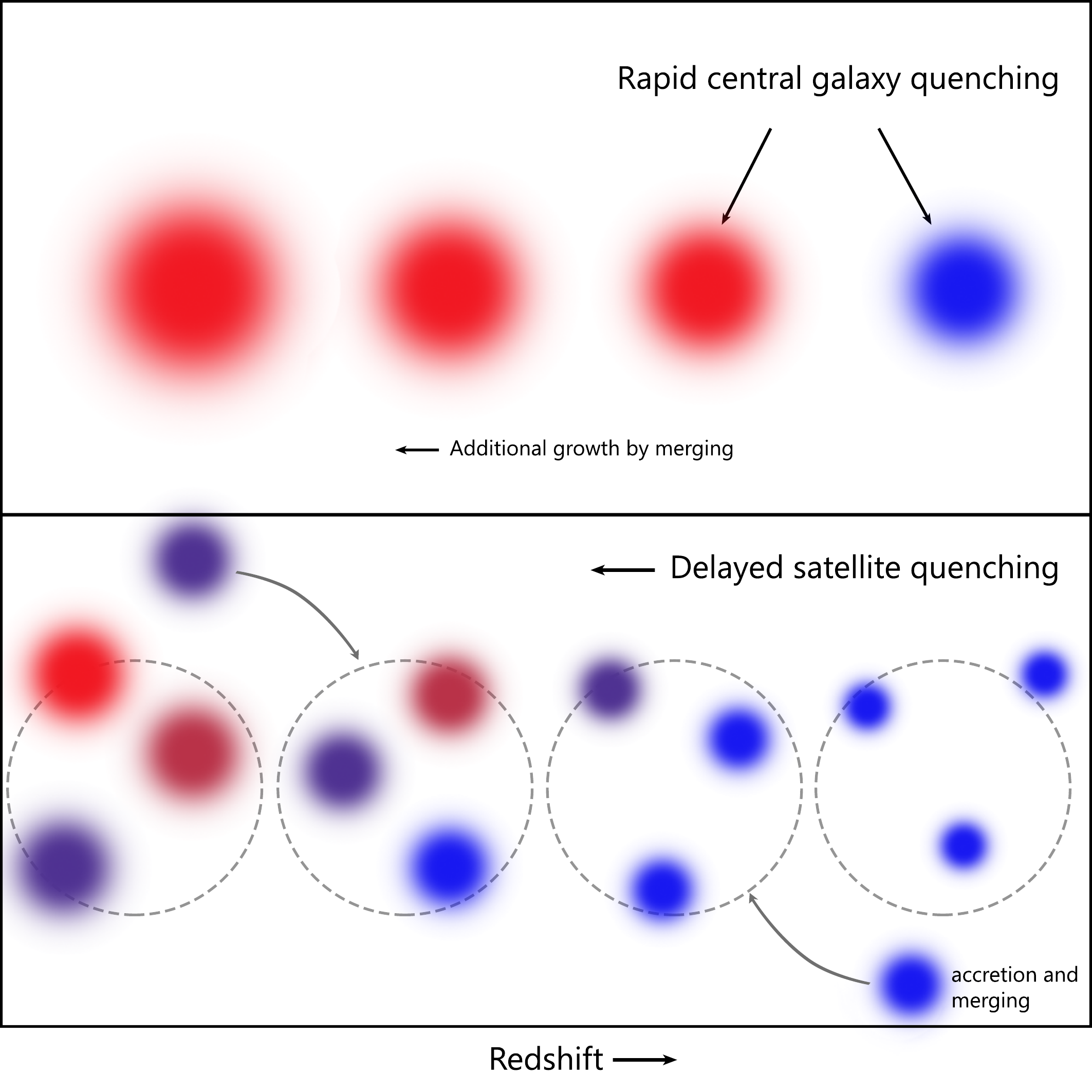}
   \caption{Top panel: star formation in central galaxies is suppressed 
     (at least in part) by halo quenching, which acts after the halo reaches 
     a critical mass of a few times $10^{12}M\solar$.
     In halos more massive than the critical mass, strong shocks are created 
     at the group virial radius and inhibit subsequent star formation by
     heating infalling gas.
     Mass and size growth continue in quenched galaxies through mergers.
     Bottom panel: satellite galaxies in the same groups experience 
     environmental quenching which starts a few Gyr after their centrals 
     quench (see Figure \ref{fig:nsf_frac}).
     Environmental processes, such as gas and stellar stripping, harassment 
     and strangulation, assist in shutting off star formation in satellite
     galaxies.
     While the quenching time scale of any individual satellite depends on its 
     orbit and on the interactions it experiences, the average quiescent 
     fraction of all satellites increases with time compared to that of 
     background galaxies at the same mass.
   }
   \label{fig:sum}
 \end{figure}

 \subsection{Sources of Uncertainty}
 \label{sec:uncert}
  The results that are presented in this study are subject to potential
  biases which may have been introduced by our galaxy selection approach.
  For example, we select satellite galaxies within a fixed mass threshold from
  their group centrals.
  Since we do not directly measure how the mass of satellite galaxies evolves
  over the studied redshift range, we cannot predict the corresponding actual
  evolution in the mass ratio value.
  An alternative selection which includes all satellite galaxies above a fixed 
  threshold mass, would imply an increasing mass ratio with time and a large 
  number of low mass satellites at low redshift.
  Consequently, the contribution of lower mass satellites may affect the 
  measured quiescent fractions of satellite and background galaxies alike
  and may change the rate of their evolution over time.
  We discuss this selection bias further in 
  Appendix \hyperref[sec:appendixa]{C}.

  An additional assumption that may affect our measurements if sufficiently 
  inaccurate is that of a constant population of satellite galaxies in the 
  studied halos. 
  If a large fraction of satellites at any given redshift are subject to 
  accretion and merging, our measurement of satellite quenching may not be 
  dominated by environmental processes as satellites do not spend enough time
  in the vicinity of the group for such processes to take effect.
  However, we note that recent studies find that most satellite galaxies orbit
  the halo on eccentric orbits and that they spend a large fraction of their 
  time outside of the group halo virial radius (but are still affected by the 
  group; e.g., Tal et al. \citeyear{tal_galaxy_2013}; Wetzel et al. 
  \citeyear{wetzel_galaxy_2013}).
  This implies that even in the presence of accretion and merging, the overall
  population of satellite galaxies may be mostly stable over time.

  Finally, the selection of constant number density of central galaxies may
  have affected the measured quiescent galaxy fractions.
  Even under the assumption that overall, a fixed cumulative number density 
  selection truly follows a given population of galaxies over time, this 
  technique may inadequately follow populations of only quiescent or star 
  forming galaxies.
  If instead we effectively follow vastly different selections of galaxy
  populations, our interpretation may suffer from significant biases.
  Nevertheless, since cumulative number density is probably our current
  best approach for tracing galaxies over different epochs, we utilize it 
  throughout this study.

\section{Summary}
 Galaxy environments have long been postulated to play a crucial role in 
 regulating the star formation rates of their resident galaxies.
 Here we studied the link between environment and star formation in groups 
 by analyzing the fraction of quiescent central and satellite galaxies 
 in the redshift range $0<z<2.5$.
 We followed central galaxy populations at fixed values of cumulative number 
 density and traced group satellites with mass ratios as low as 1:10.
 In addition, we analyzed background galaxies (which act as centrals in their
 own host halos) at the same mass and redshift as the satellites.
 Central galaxies in our sample span a wide range of stellar masses 
 ($3.5\pow10{9} < M/M\solar < 2.2\pow10{11}$) over seven redshift bins and
 four number density selection values.
 We found clear evidence for star formation suppression in group halos and
 measured different quenching times for satellite and 
 central galaxies.

  \begin{table*}[t]

    \caption{Central galaxy selection criteria for all number density samples}
    \centering
    \begin{tabular}{c c c c c c c}
      \hline\hline
      Selection & Data set & Redshift & Central mass range &
      $M_{\mathrm{med}}$ & R$_{\mathrm{vir}}$ & N$_\mathrm{cen}$ \\
      & & & & $(M\solar)$ & (kpc) & \\

      \hline {\vspace{-5px}}\\

      & {\vspace{5px}}SDSS & $0.02<z<0.04$ & $10.70<$log$(M/M\solar)<11.00$ &
      6.2$\pow10{10}$ & 220 & 667 \\
      $n=6\pow10{-4}$ Mpc$^{-3}$ &  
        \multirow{4}{*}{UltraVISTA} & $0.2<z<0.4$ & 
        $10.54<$log$(M/M\solar)<10.85$ & 4.9$\pow10{10}$ & 
        200 & 301 \\
      & & $0.4<z<0.6$ & $10.45<$log$(M/M\solar)<10.75$ & 4.0$\pow10{10}$ & 
        180 & 531 \\
      & & $0.6<z<0.85$ & $10.35<$log$(M/M\solar)<10.65$ & 3.1$\pow10{10}$ & 
        150 & 1251 \\
      $\log(M_\star/M\solar)=-0.07z^2-0.37z+10.81$ & {\vspace{5px}} & 
        $0.85<z<1.2$ & $10.20<$log$(M/M\solar)<10.50$ & 2.2$\pow10{10}$ & 
        110 & 2858\\
      & \multirow{3}{*}{3D-HST} & $1.2<z<1.8$ & $9.93<$log$(M/M\solar)<10.23$ & 
        1.2$\pow10{10}$ & 85 & 970 \\
      & & $1.8<z<2.5$ & $9.55<$log$(M/M\solar)<9.85$ & 5.0$\pow10{9}$ & 
        60 & 1827 \\
      & {\vspace{3px}}& $2.5<z<3.5$ & $8.87<$log$(M/M\solar)<9.17$ & 
        1.0$\pow10{9}$ & 30 & 3530 \\

      \hline {\vspace{-5px}}\\

      & {\vspace{5px}}SDSS & $0.02<z<0.04$ & $10.83<$log$(M/M\solar)<11.13$ & 
      7.9$\pow10{10}$ & 250 & 251 \\
      $n=4\pow10{-4}$ Mpc$^{-3}$ &  
        \multirow{4}{*}{UltraVISTA} & $0.2<z<0.4$ & 
        $10.69<$log$(M/M\solar)<10.99$ & 6.9$\pow10{10}$ & 220 & 261 \\
      & & $0.4<z<0.6$ & $10.64<$log$(M/M\solar)<10.94$ & 6.1$\pow10{10}$ & 
        200 & 460 \\
      & & $0.6<z<0.85$ & $10.57<$log$(M/M\solar)<10.87$ & 5.2$\pow10{10}$ & 
        180 & 1080 \\
      $\log(M_\star/M\solar)=-0.10z^2-0.18z+10.90$ & {\vspace{5px}} & 
        $0.85<z<1.2$ & $10.45<$log$(M/M\solar)<10.75$ & 4.1$\pow10{10}$ & 
        150 & 2281\\
      & \multirow{3}{*}{3D-HST} & $1.2<z<1.8$ & 
        $10.25<$log$(M/M\solar)<10.55$ & 2.5$\pow10{10}$ & 110 & 707 \\
      & & $1.8<z<2.5$ & $9.92<$log$(M/M\solar)<10.22$ & 
        1.2$\pow10{10}$ & 75 & 915 \\
      & {\vspace{3px}}& $2.5<z<3.5$ & $9.29<$log$(M/M\solar)<9.59$ & 
        2.8$\pow10{9}$ & 40 & 3054\\

      \hline {\vspace{-5px}}\\

      & {\vspace{5px}}SDSS & $0.02<z<0.04$ & $10.97<$log$(M/M\solar)<11.27$ & 
      1.1$\pow10{11}$ & 350 & 50 \\
      $n=2\pow10{-4}$ Mpc$^{-3}$ &  
        \multirow{4}{*}{UltraVISTA} & $0.2<z<0.4$ & 
        $10.89<$log$(M/M\solar)<11.19$ & 1.0$\pow10{11}$ & 300 & 165 \\
      & & $0.4<z<0.6$ & $10.87<$log$(M/M\solar)<11.17$ & 
        9.8$\pow10{10}$ & 270 & 250 \\
      & & $0.6<z<0.85$ & $10.83<$log$(M/M\solar)<11.13$ & 
        9.0$\pow10{10}$ & 240 & 610 \\
      $\log(M_\star/M\solar)=-0.10z^2-0.04z+11.03$ & {\vspace{5px}} & 
        $0.85<z<1.2$ & $10.76<$log$(M/M\solar)<11.06$ & 
        7.7$\pow10{10}$ & 200 & 1294\\
      & \multirow{3}{*}{3D-HST} & $1.2<z<1.8$ & 
        $10.61<$log$(M/M\solar)<10.91$ & 5.6$\pow10{10}$ & 150 & 526 \\
      & & $1.8<z<2.5$ & $10.35<$log$(M/M\solar)<10.65$ & 
        3.2$\pow10{10}$ & 100 & 595 \\
      & {\vspace{3px}}& $2.5<z<3.5$ & $9.86<$log$(M/M\solar)<10.16$ & 
        9.9$\pow10{9}$ & 60 & 1601\\

      \hline {\vspace{-5px}}\\

      & {\vspace{5px}}SDSS & $0.02<z<0.04$ & $11.05<$log$(M/M\solar)<11.35$ & 
      1.5$\pow10{11}$ & 380 & 21 \\
      $n=1\pow10{-4}$ Mpc$^{-3}$ &  
        \multirow{4}{*}{UltraVISTA} & $0.2<z<0.4$ & 
        $11.02<$log$(M/M\solar)<11.32$ & 1.4$\pow10{11}$ & 350 & 133 \\
      & & $0.4<z<0.6$ & $11.00<$log$(M/M\solar)<11.30$ & 
        1.4$\pow10{11}$ & 330 & 167 \\
      & & $0.6<z<0.85$ & $10.99<$log$(M/M\solar)<11.29$ & 
        1.3$\pow10{11}$ & 310 & 369 \\
      $\log(M_\star/M\solar)=-0.08z^2-0.02z+11.17$ & {\vspace{5px}} & 
        $0.85<z<1.2$ & $10.96<$log$(M/M\solar)<11.26$ & 
        1.2$\pow10{11}$ & 250 & 714\\
      & \multirow{3}{*}{3D-HST} & $1.2<z<1.8$ & 
        $10.85<$log$(M/M\solar)<11.15$ & 9.2$\pow10{10}$ & 190 & 256 \\
      & & $1.8<z<2.5$ & $10.65<$log$(M/M\solar)<10.95$ & 
        6.1$\pow10{10}$ & 140 & 437 \\
      & {\vspace{3px}}& $2.5<z<3.5$ & $10.27<$log$(M/M\solar)<10.57$ & 
        2.6$\pow10{10}$ & 70 & 759\\

      \hline

      \label{tab:selection2}
    \end{tabular}
  \end{table*}

 Initial star formation suppression in central galaxies takes place early.
 Figure \ref{fig:nsf_n} shows that at $z\sim2.5$, at least 30\% of the 
 centrals in each of the two most massive samples are already quenched.
 Even the lowest mass centrals in our sample had already started quenching
 by $z\sim1.5$.
 We utilized these measurements to calculate a threshold halo mass for 
 star formation suppression and found it to be in the range
 $10^{12}<M_{crit}/M\solar<10^{13}$ (Figure \ref{fig:nsf_frac}).
 This value is relatively constant at $1.3<z<3.0$ and it is consistent with 
 theoretical predictions from the halo quenching model.
 Local processes, which were not considered in this analysis, likely contribute
 to quenching as well.

 Satellite galaxies begin quenching a few Gyr after the onset of star formation
 cessation in the centrals.
 Figure \ref{fig:nsf_frac} shows that at $z\sim1.5$ the average fraction of 
 quiescent satellite and background galaxies is similar to one another but 
 significantly lower than that of centrals.
 At that redshift, centrals had already been quenching for 1-3 Gyr.
 At $z\sim0$, the fraction of quiescent satellites 
 approaches and nearly reaches the fraction of quiescent centrals.
 This delayed quenching was also reported by Wetzel et al. 
 (\citeyear{wetzel_galaxy_2012}) in low redshift groups.

 Additional evidence for the difference between central and satellite galaxy
 quenching comes from the different quiescent fractions of satellite and 
 background galaxies at $z<1.5$.
 Since sources that were selected as background galaxies act as centrals in 
 their respective halos, the black and blue lines in Figure \ref{fig:nsf_frac} 
 represent the difference between central and satellite quenching at the same
 mass and redshift.
 This is in agreement with studies who find that satellite galaxies are on 
 average more quenched than same-mass centrals (e.g., Wetzel et al. 
 \citeyear{wetzel_galaxy_2012}).
 Overall, the observed quenching in satellites is consistent 
 with environmental processes (mainly gas heating and stripping) which act to 
 suppress star formation in galaxy group halos.


 In summary, we find that while star formation gets suppressed in all galaxies
 in groups, the processes that govern this suppression are not universal 
 (Figure \ref{fig:sum}).
 Our analysis shows that central galaxies in groups undergo quenching 
 which is consistent with the halo quenching model while their satellites 
 experience additional quenching by environmental processes.
 The onset of star formation suppression also varies between centrals 
 and satellites with early central quenching and delayed quenching of 
 satellites.

 \begin{figure*}
   \includegraphics[width=1.0\textwidth]{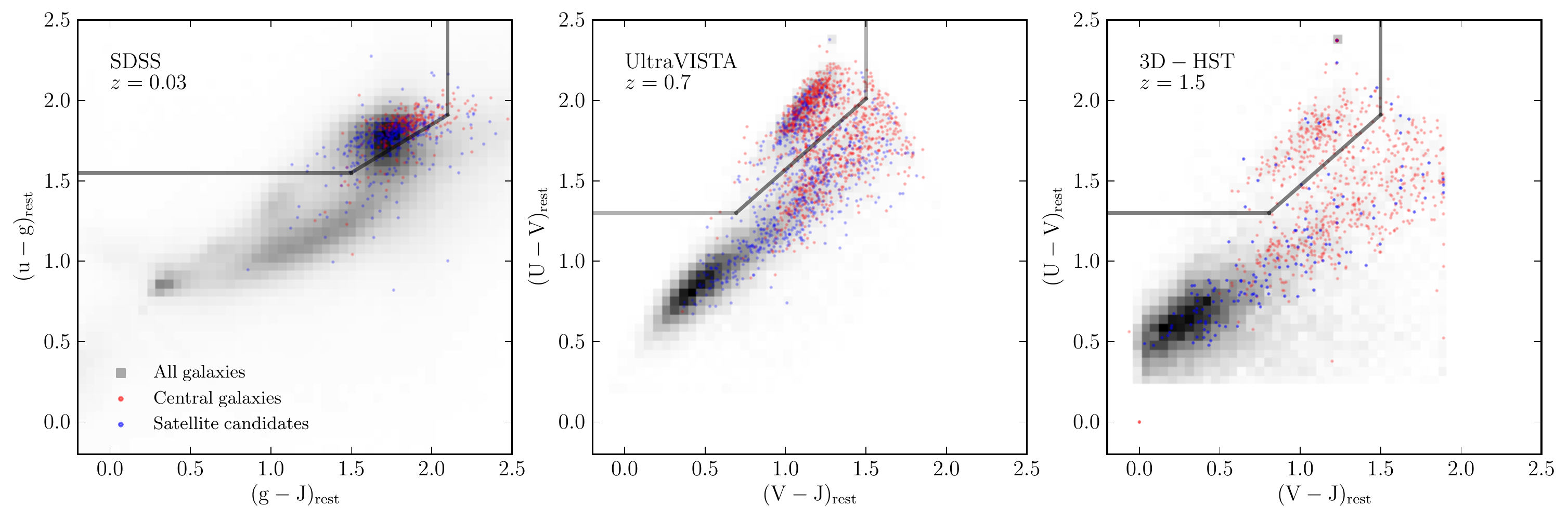}
   \caption{Rest frame color based selection of quiescent galaxies from SDSS, 
     UltraVISTA and 3D-HST/CANDELS.
     Gray points represent the distribution of all galaxies in the catalogs 
     while red and blue points show selected central galaxies and satellite 
     candidates, respectively.
     At $z>0.2$ quiescent galaxies were selected from a rest frame $UVJ$ 
     diagram according to the thresholds found by Muzzin et al. 
     (\citeyear{muzzin_evolution_2013}).
     Lower redshift galaxies were selected in rest frame $ugJ$ space to roughly
     match the high redshift selection.
     The enhanced vertical scatter in the SDSS points largely results from
     relatively poor $u-$band photometry.
   }
   \label{fig:uvj}
 \end{figure*}

\begin{acknowledgements}
%
  We thank Guillermo Barro and Alis Deason for engaging discussions and for 
  commenting on an early draft of the paper.

  TT is supported by an NSF Astronomy and Astrophysics Postdoctoral Fellowship 
  under award AST-1202667.

  This study is based in part on a K$_{s}$-selected catalog of the 
  COSMOS/UltraVISTA field from Muzzin et al. (\citeyear{muzzin_public_2013}).  
  The catalog contains PSF-matched photometry in 30 photometric bands covering 
  the wavelength range 0.15$\micron$ $\rightarrow$ 24$\micron$ and includes the 
  available $GALEX$ (Martin et al. \citeyear{martin_galaxy_2005}), 
  CFHT/Subaru (Capak et al. \citeyear{capak_first_2007}), 
  UltraVISTA (McCracken et al. \citeyear{mccracken_ultravista:_2012}), 
  S-COSMOS (Sanders et al. \citeyear{sanders_s-cosmos:_2007}), and 
  zCOSMOS (Lilly et al. \citeyear{lilly_zcosmos_2009}) datasets.
  The catalog was derived using data products from observations made with ESO 
  telescopes at the La Silla Paranal Observatory under ESO programme 
  ID 179.A-2005 and on data products produced by TERAPIX and the Cambridge 
  Astronomy Survey Unit on behalf of the UltraVISTA consortium.

  This work is also based in part on observations taken by the 3D-HST Treasury 
  Program (GO 12177 and 12328) with the NASA/ESA HST, which is operated by the 
  Association of Universities for Research in Astronomy, Inc., under NASA 
  contract NAS5-26555.

  Funding for the SDSS and SDSS-II has been provided by the Alfred P. 
  Sloan Foundation, the Participating Institutions, the National Science 
  Foundation, the U.S. Department of Energy, the National Aeronautics and
  Space Administration, the Japanese Monbukagakusho, the Max Planck Society, 
  and the Higher Education Funding Council for England. The SDSS Web Site is
  http://www.sdss.org/.
  The SDSS is managed by the Astrophysical Research Consortium for the 
  Participating Institutions. The Participating Institutions are the 
  American Museum of Natural History, Astrophysical Institute Potsdam, 
  University of Basel, University of Cambridge, Case Western Reserve 
  University, University of Chicago, Drexel University, Fermilab, the 
  Institute for Advanced Study, the Japan Participation Group, Johns 
  Hopkins University, the Joint Institute for Nuclear Astrophysics, the 
  Kavli Institute for Particle Astrophysics and Cosmology, the Korean 
  Scientist Group, the Chinese Academy of Sciences (LAMOST), Los Alamos 
  National Laboratory, the Max-Planck-Institute for Astronomy (MPIA), the 
  Max-Planck-Institute for Astrophysics (MPA), New Mexico State University, 
  Ohio State University, University of Pittsburgh, University of Portsmouth, 
  Princeton University, the United States Naval Observatory, and the 
  University of Washington.
\end{acknowledgements}

\section*{Appendix A}
\section*{cumulative fixed number density selection parameters}
 \label{sec:appendixa}
 Central galaxy selection criteria are presented in Table \ref{tab:selection2}.
 The selection column shows the fixed cumulative number density value of each 
 sample, as well as its derived mass-redshift relation (Section 
 \ref{sec:nselect}).
 Stellar mass ranges were calculated such that each log($M/M\solar$) bin of 
 size 0.3 dex had a median central galaxy mass according to its respective
 mass-redshift relation.
 Virial radius estimates were calculated using catalogs based on the 
 semi-analytic models of Guo et al. \citeyear{guo_dwarf_2011} for galaxies
 at the same redshift and mass.
 The number of central galaxies in each bin are given in the final column of 
 Table \ref{tab:selection2}.
 
\section*{Appendix B}
\section*{$UVJ$ and \lowercase{ug}J selection}
 \label{sec:appendixb}
 Quiescent galaxies at $z>0.2$ were selected from the UltraVISTA and 
 3D-HST/CANDELS catalogs based on their location on a rest frame $UVJ$ diagram.
 The ancillary data sets and SED modeling techniques that were used to derive 
 these colors for both catalogs are similar and therefore so are the 
 galaxy selection limits that were utilized in this study.
 The middle and right-hand panels of Figure \ref{fig:uvj} show the distribution
 of galaxy colors in the catalogs, as well as the selection thresholds 
 (black lines) which were adopted from Muzzin et al. 
 (\citeyear{muzzin_evolution_2013}).
 
 At $z\sim0$, rest frame color estimates require very small corrections 
 from observed colors.
 More specifically, $u-g$ color values in the NYU-VAG catalog
 were derived relying heavily on $u-$band photometry from SDSS.
 The response function of the SDSS photometric system in this filter band 
 is notoriously poor and as a consequence flux estimates suffer from large 
 errors (Fukugita et al. \citeyear{fukugita_sloan_1996}).
 While the $u-g$ and $U-V$ colors that were used in this study are very 
 similar in wavelength coverage, the enhanced scatter in $u-$band photometry
 implies that the $u-g$ measurement is more uncertain.
 The left panel of Figure \ref{fig:uvj} shows the distribution of galaxy colors
 in SDSS. 
 It is evident from the figure that the division between star forming and 
 quiescent galaxies is far from perfect as the two populations are somewhat
 mixed.
 
 In order to match the selection in $ugJ$ to that of $UVJ$ we followed 
 Williams et al. (\citeyear{williams_detection_2009}) and 
 Muzzin et al. (\citeyear{muzzin_evolution_2013}) and tried to contain as 
 much of the quiescent galaxy peak while minimizing contamination from star 
 forming galaxies.
 The lines in the left hand panel of Figure \ref{fig:uvj} show these selection
 thresholds.
 Despite significant scatter, the majority of all quiescent central and 
 satellite galaxies are likely included in our selection.

\section*{Appendix C}
\section*{Constant mass selection of central and satellite galaxies}
\label{sec:appendixc}
\begin{figure}
   \includegraphics[width=0.48\textwidth]{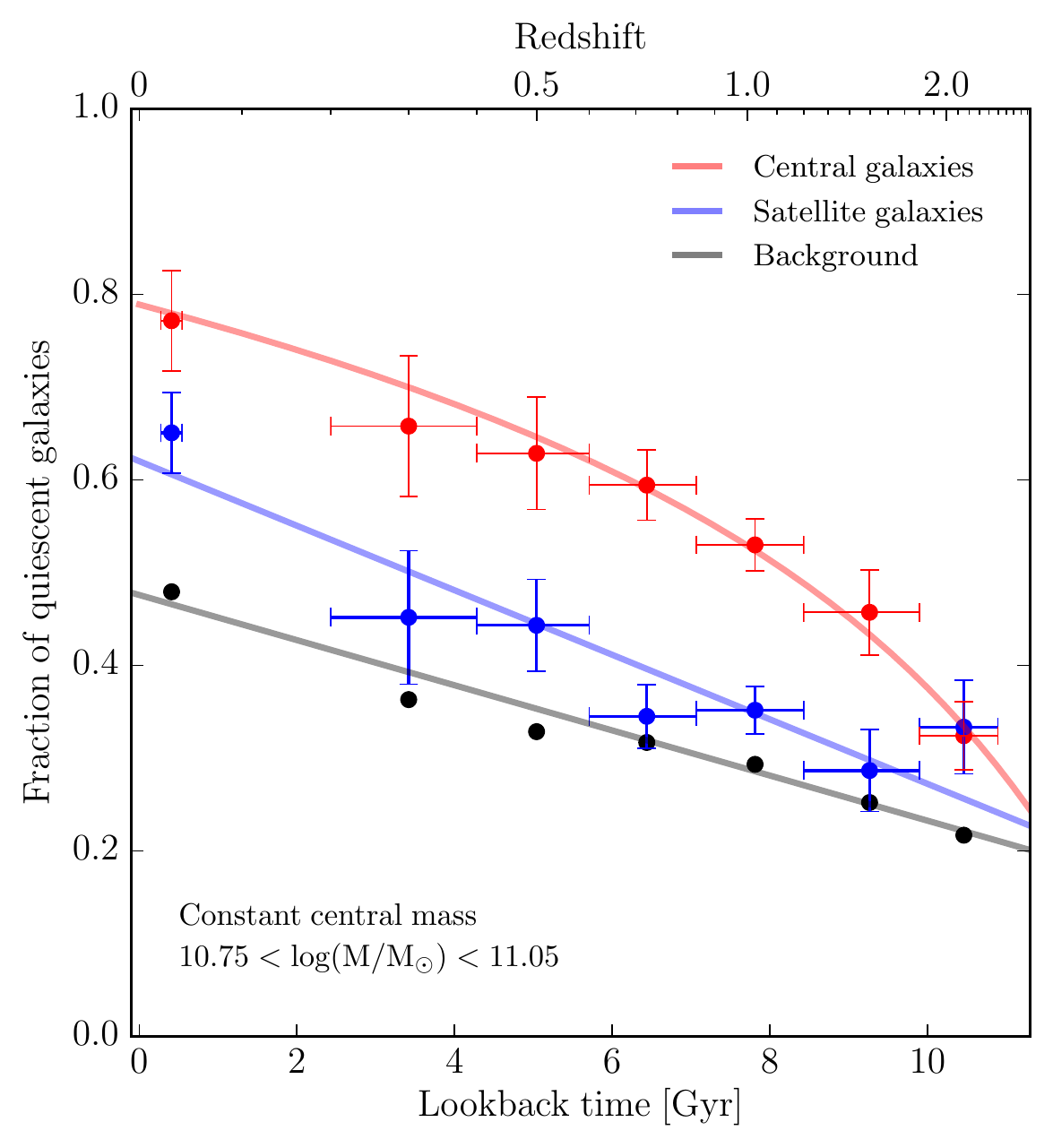}
   \caption{Evolving quiescent fractions of central, satellite and 
     background galaxies, selected within a constant stellar mass bin.
     Central galaxies at all redshifts have stellar mass estimates of
     $10.75<\log(M/M\solar)<11.05$ while all satellite and background galaxies
     are more massive than $\log(M/M\solar)>9.75$.
     Even at a constant mass selection, satellite galaxies begin a delayed 
     quenching after their respective centrals, and at a faster rate than 
     background galaxies.
   }
   \label{fig:mconst}
 \end{figure}

 One potential bias that could affect our analysis of the evolving quenched 
 fractions of satellite galaxies (Section \ref{sec:sats}) stems from the 
 number density selection approach that we utilized in this study.
 Since we do not directly follow the mass growth of satellite galaxies but 
 rather require a fixed ratio between central and satellite masses,
 some of the implied evolution of the quiescent fractions may be the result 
 of mass growth over the same redshift range.
 Here we test the robustness of our results by repeating the analysis for 
 a constant mass selected sample of central and satellite galaxies.
 Central galaxies at all redshifts are in the mass range 
 $10.75<\log(M/M\solar)<11.05$ while all satellite and background galaxies
 are more massive than $\log(M/M\solar)>9.75$.
 Figure \ref{fig:mconst} demonstrates that even in the absence of any mass 
 evolution, satellite galaxies experience delayed quenching after their 
 centrals, at a more rapid rate than background galaxies at the same mass
 and redshift.
 We conclude that the observed trends that are presented in this study do
 not entirely result from the mass evolution of satellite galaxies.
 
\bibliographystyle{yahapj}
\bibliography{ms}

\end{document}